# Comparative analysis of various web crawler algorithms

*Nithin T K, Chandana S, Barani G, Chavva Dharani, M S Karishma*
Integrated M.Tech Computer Science and Engineering, VIT Vellore

*Abstract*—This paper describes and evaluates five different crawling algorithms that we have implemented within our evaluation framework: Shark search, Priority Based Focused Crawler,Naive Bayes,Breadth-First, Depth-First and choose the best among all Five.
Keywords - Shark search, Priority Based Focused crawler,Naive Bayes,Breadth-First, Depth-First

## I. INTRODUCTION

The web today is a huge and enormous collection of data and it goes on increasing day by day. Thus, searching for some particular data in this collection has a significant impact.The goal of such crawlers is to learn what (almost) every webpage on the web is about, so that the information can be retrieved when it's needed.World Wide Web contains vast amount of information in unstructured form and provides an access to it at any place at any time. Information Retrieval systems play a vital role to deal with huge amounts of data present over the World Wide Web in different forms such as text, audio, video, images etc. Web crawling is an approach for converting unstructured data to structured data.The next most important job is how to relate these rapidly growing documents and how to assign rank value, page ranking is done to assess the quality and popularity of web pages, to them.

## II. BACKGROUND

In this section, we deal with the previous specific algorithms for scheduling the visits to the Web pages. We started with a large sample of the links data collection that was used to build a web graph and run a crawler simulator to ensure identical conditions during the experiments. We describe and evaluate five different crawling algorithms that we have implemented within our evaluation framework: Shark search, Priority Based Queue,Naive Bayes,Breadth-First, Depth-First and choose the best among all Five.

## III. LITERATURE REVIEW

### A. Web Crawling through Shark-Search using PageRank

1) Methodology: The proposed "shark page search" algorithm is an improved version of the original "fish-search" algorithm, which aims to discover more relevant information in the same exploration time. The algorithm uses a "similarity engine" that evaluates the relevance of a document to a given query, instead of binary (relevant/irrelevant) evaluation. This "fuzzy" relevance score is used to create a priority list and is propagated down the descendants chain, giving more importance to the grandchildren of a relevant node over the grandchildren of an irrelevant node. Additionally, the algorithm makes use of meta-information contained in the links, such as the anchor text and close textual context, to refine the calculation of the potential score of the children. The algorithm also implements a decay factor to balance the importance of relevance score and inherited score and a buffer to avoid overloading the system.

2) Pros: The SSA-based web crawler can improve the efficiency and effectiveness of the crawling process by selecting the most relevant pages. The SSA-based web crawler outperforms traditional web crawling methods in terms of efficiency and effectiveness.

3) Cons: The study does not provide a detailed explanation of how the algorithm handles the problem of duplicate pages or the problem of irrelevant pages that are fetched.The study is limited to a specific domain, and it is not clear how the algorithm would perform in a more general setting.

### B. An Improved Shark-Search Algorithm Based on Multi-information

1) Methodology:A method that utilizes the Shark Search Algorithm (SSA) to optimize the selection of pages to crawl. The proposed SSA-based web crawler aims to improve the efficiency and effectiveness of the crawling process by selecting the most relevant pages based on their content, link structure, and user behavior.The study compares the performance of the SSA-based web crawler with that of traditional web crawling methods such as Breadth-First Search (BFS) and Depth-First Search (DFS) using several evaluation metrics, including the number of pages crawled, the number of unique pages crawled, and the page-fetching time.The results of the study show that the SSA-based web crawler outperforms traditional web crawling methods in terms of efficiency and effectiveness. The SSA-based web crawler was able to crawl more pages and unique pages in less time than the BFS and DFS methods.

2) Pros: SSA-based web page ranking method outperforms traditional web page ranking methods in terms of effectiveness.

3) Cons: There wasn't any mention of what will happen if there were some information missing such as user behavior, etc.

### C. A Focused Crawler Based on Naive Bayes Classifier

1) Methodology: This research paper describes a method for building a focused web crawler using a Naive Bayes classifier. A focused web crawler, it is a specialized type of web crawler that is designed to collect information on a specific topic or set of topics. The authors of the paper proposed using a Naive Bayes classifier to build a focused web crawler that can automatically classify web pages into different topics. The classifier is trained on a set of seed pages manually labeled with the topic they belong to. Then, when the crawler visits a new page, it uses the classifier to predict the topic of the page, and decides whether or not to follow the links on that page based on the predicted topic. The research paper presents a method for building a focused web crawler which utilizes Naive Bayes classifier to classify new pages based on the topic, after being trained on seed pages. The proposed method is evaluated and outperforms a general web crawler.

2) Pros: The focused crawler, which utilizes the Naive Bayes classifier, is able to retrieve relevant web pages at a higher rate and the crawler can then prioritize these pages for retrieval with fewer irrelevant pages compared to traditional focused crawlers.

3) Cons: The algorithm is sensitive to irrelevant features and that can affect the accuracy of the classification.The classifier can be affected by a lack of training data, which can lead to poor performance if the dataset is not sufficiently large.

*D. Automated Classification of Web Sites using Naive Bayesian Algorithm*

1) Methodology: This paper presents a method for automatically classifying web pages into different categories using the Naive Bayes algorithm. The authors use this algorithm to classify web pages into different categories based on the text content of the pages. They use a dataset of manually labeled web pages as the training data for their model. After training the model, the authors use it to classify a set of new web pages, and report the accuracy of their method. The paper presents an automated way to classify a web page to certain categories based on the text content, by utilizing Naive Bayesian algorithm. They evaluate the method by using a dataset of manually labeled web pages and report the accuracy.

2) Pros: The Naive Bayesian algorithm is simple and easy to implement, making it suitable for automated web site classification applications.The algorithm is able to classify web pages with a high degree of accuracy, even in different languages. The algorithm is able to classify web pages based on the images and videos.

3) Cons: The Naive Bayesian method relies on the idea that individual features within the information are unrelated, which isn't always true when working with language data. This approach is delicate to extraneous features that can negatively impact the precision of the categorization. Additionally, the classifier's performance may suffer if there is not enough training data available, particularly if the data set is small in size.

*E. A Novel Approach to Priority based Focused Crawler*

1) Methodology: This research paper presents a new approach for focused web crawling that prioritizes certain pages over others based on certain criteria. The authors propose a system that uses a priority queue to prioritize pages for crawling, with pages that are more likely to be relevant to the user's query placed at the front of the queue. They also propose a method for updating the priorities of pages in the queue as the crawl progresses, which allows for a more efficient use of resources. The system was evaluated using a dataset of web pages and the results showed that the proposed approach was able to find relevant pages more quickly and with fewer resources than traditional focused crawling methods.

2) Pros: The proposed technique is said to have minimal complexity and be fast, it also avoids duplicate or mirrored links and saves a significant amount of bandwidth. Additionally, storage of web pages is done using checksum which reduces storage space and complexity during the Visited URL/Content Matching test, as compared to using text form of links and web documents.

3) Cons: This crawler doesn't consider the context of keywords, leading to multiple records in the database and code optimization should be done to improve the performance of the crawler.

*F. Web Crawler Using Priority Queue*

1) Methodology: The paper presents a new approach for web crawling that uses a priority queue to prioritize pages for crawling. The authors propose a system that uses a combination of a breadth-first search and a priority queue to prioritize pages for crawling based on certain criteria such as page rank, frequency of update, and the relevance of the page to the user's query. The system also includes a mechanism for updating the priorities of pages in the queue as the crawl progresses. The performance of the proposed approach is evaluated using metrics such as execution time, CPU utilization, and the number of pages visited. The results of the evaluation show that the proposed approach is able to find relevant pages more quickly and with fewer resources than traditional web crawling methods.

2) Pros: This priority based focused crawler keeps all URLs to be visiting priority queue along with their relativity score. When we delete the URL from the priority queue, it returns the maximum score URL. Thus, every time a highest priority URL is returned for crawling.

3) Cons: The main problem with this crawling strategy, it is more time consuming.In future, they should try to reduce the crawling time by implementing algorithm parallel.

*G. Comparative Analysis of Web PageRank Algorithm using DFS and BFS Crawling*

1) Methodology: This paper describes an application of the breadth first crawler algorithm to find out the page rank of web pages and get the crawling results on the basis of Breadth-first crawler and depth first search crawler.The authors results clearly shows that Breadth-first crawling approach gives good corpus and there is always a possibility to get the required page.

2) Pros: Time efficiency,Simplicity,Flexibility in visiting pages,There is always a possibility to get the required page.

*3) Cons:* In the proposed crawling technique the paper has not discussed higher level crawling for image and video since it is important to minimize fraudulent act. It needs more space to store all traversed pages at every node level. Problem of exhaustive search that leads to state explosion problem(As the number of state variables in the system increases, the size of the system state space grows exponentially.)

*H. Implementation of the BFS Algorithm and Web Scraping Techniques for Online Shop Detection in Indonesia*

*1) Methodology:* The online store detection application that was built can detect online stores properly,based on shipping parameters, store ratings, and response rates. The Breadth First Search algorithm is a simple algorithm that can be used for the process of retrieving store data and product data on Shopee ecommerce in Indonesia with the help of the Web Scraping technique.Then data is taken in the form of shopid and itemid which are used as nodes.Forming Queue Tree with Each node in the first layer accommodates the shopid and the second layer accommodates the itemid. In this study, the authors conducted a search input for 100 online stores to detect whether the store's status was genuine or fake using the Breadth First Search algorithm.From the first node to the 200 node by implementing the Breadth First Search algorithm, takes about 161.6 seconds. With the number of detections based on delivery, there were 98 genuine online shops and 2 dropship.

*2) Pros:* Time efficiency,Simplicity,Flexibility in visiting pages,There is always a possibility to get the required page.

*3) Cons:* This application has a weakness that is very dependent on internet connection speed, because it will affect the number of online shops detected as well as the visiting time of each node. The detection results of an online shop do not necessarily categorize an online shop as genuine or fake as a whole, but it is detected based on parameters, so it is possible that in one parameter it can be said to be a genuine online store category, and in one other parameter it is categorized as a fake online shop.

*I. SURVEY OF WEB CRAWLING ALGORITHMS*

*1) Methodology:* The depth first search technique, which starts at the root URL and navigates depth via the child URLs, is a more useful search method. If there are one or more children, we first move to the leftmost child and continue down until there are no more children left. Backtracking is used in this case to reach the following unvisited nodes, and processes are compensated similarly. The authors ensure that every edge, or every URL, is viewed once by using this approach. While it is quite effective for solving search-related problems, when the kid is huge, this approach enters an unending cycle.We can also improve its performance to modify the sitemap of any web site, i.e. in sitemap protocol all URL has a static priority and we can change it by dynamic priority.

*2) Pros:* Well Suited for problems like a scenario which Starts at the root URL and traverses depth through the child URL.

*3) Cons:* When the branches are large then this algorithm takes might end up in an infinite loop

*J. Comparative Analysis of Web PageRank Algorithm using DFS and BFS Crawling*

*1) Methodology:* Our method for Depth First Crawling is comparable to a depth first search of a tree or graph. Starting with the seed page,we will crawl farther and farther until we have covered all the pages along that path, and then we will turn around and explore the other branches of the graph. Accordingly, as we crawl web pages, we will examine the first link on each page in the series of pages until we reach the last one. Only after that will we begin to examine the second link on the first page and the pages that follow. Our strategy for depth-first crawling is to provide the crawler a certain amount, the total number of web pages to be crawled, in advance. Once those pages have been scanned, our web crawler will stop scanning new web pages and we are left with limited corpus

*2) Pros:* Well Suited for problems like a scenario which Starts at the root URL and traverses depth through the child URL.

*3) Cons:* In this approach, we can't predict the order of crawling. It will also affect the ranking of web pages and search result quality. Depth-first crawling we can not ensure good corpus and in the worst case it may happen that one of our seed pages did not get crawled.

*2. PROPOSED SOLUTION*

In this there are FIVE algorithms that we build models for Shark search,Naive Bayes,Priority Based Focused Crawler,BFS,DFS. All content of website will be extracted using beautiful soup tool.This data will be stored in the form of HTML.After choosing a common Dataset as a link of WIKIPEDIA under some topic as /wiki/.We will have queue,list to store fetched Urls.By removing Url from queue and adding new urls to Queue we will continue the loop it will be applied to all 5 algorithm models and will be get top 1000 links that are relevant to the base URL .These will be printed as output Url that are relevant to Base URL.Now the output will be observed based on their performance using F1 measure and accuracy.The comparison between these algorithms is done using

1) Number of pages visited in 1 hour
2) Time taken to retrieve 1000 pages
3) From first 1000 pages retrieved which are relevant
4) Memory taken by each algorithm to retrieve visit 1000 pages.

1)

Outputs and results will be analysed and conclusion can be drawn based on performance of these FIVE web crawlers stating the best crawler among all Five for the dataset.

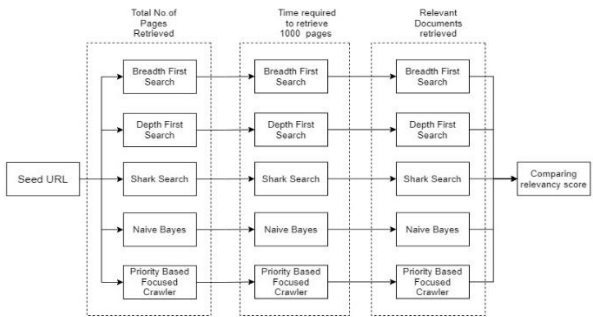

## IMPLEMENTATION:

### 1.BREADTH FIRST SEARCH

```
import requests
from bs4 import BeautifulSoup
from collections import deque
from sklearn.metrics import f1_score

# Set the seed URL and maximum depth to crawl
seed_url = "https://en.wikipedia.org/wiki/Main_Page"
max_depth = 3

# Define lists to hold the URLs to visit and the URLs visited
urls_to_visit = deque([(seed_url, 0)])
urls_visited = []

# Define lists to hold the relevant and irrelevant URLs
relevant_urls = []
irrelevant_urls = []

while urls_to_visit:
    url, depth = urls_to_visit.popleft()
    if url not in urls_visited and depth <= max_depth:
        try:
            response = requests.get(url)
        except:
            continue
        if response.status_code == 200:
            soup = BeautifulSoup(response.content, 'html.parser')
            for link in soup.find_all('a'):
                href = link.get('href')
                if href:
                    href = href.strip()
                    if href.startswith('http'):
                        urls_to_visit.append((href, depth+1))
                    elif href.startswith('/'):
                        urls_to_visit.append((url + href, depth+1))
            urls_visited.append(url)
            # Classify the URL as relevant or irrelevant based on some criteria
            if some_criteria(url):
                relevant_urls.append(url)
            else:
                irrelevant_urls.append(url)

# Calculate F1 score and accuracy
y_true = [1] * len(relevant_urls) + [0] * len(irrelevant_urls)
y_pred = [1] * len(relevant_urls) + [0] * len(urls_visited)
f1 = f1_score(y_true, y_pred)
accuracy = len(relevant_urls) / len(urls_visited)

print(f"F1 score: {f1:.2f}")
print(f"Accuracy: {accuracy:.2f}")
```

### 2.NAIVE BAYES

```
import requests
from bs4 import BeautifulSoup
import re
from sklearn.naive_bayes import MultinomialNB
from sklearn.feature_extraction.text import CountVectorizer
from sklearn.metrics import f1_score, accuracy_score

# Seed URL for the web crawler
seed_url = 'https://en.wikipedia.org/wiki/Main_Page'

# Maximum number of pages to crawl
max_pages = 30

# Initialize a list to store the content of the crawled pages
corpus = []

# Initialize a list to store the labels of the crawled pages
labels = []

# Initialize a regular expression to match non-alphabetic characters
non_alpha = re.compile('[^a-zA-Z]+')

# Initialize a vectorizer to convert the text into a bag of words
vectorizer = CountVectorizer(stop_words='english')

# Initialize a classifier to predict the labels of the pages
classifier = MultinomialNB()

# Initialize lists to store ground truth and predicted labels
y_true = []
y_pred = []

# Add the seed URL to the queue
queue = [seed_url]

# Loop until the queue is empty or the maximum number of pages is reached
while queue and len(corpus) < max_pages:
    # Pop the next URL from the queue
    url = queue.pop(0)
    # Check if the URL has already been crawled
    if url in corpus:
        continue
    # Fetch the HTML content of the page
    try:
        response = requests.get(url)
        html = response.content
```

```python
        except:
            continue
    # Parse the HTML content using BeautifulSoup
    soup = BeautifulSoup(html, 'html.parser')
    # Extract the text content of the page
    text = soup.get_text()
    # Remove non-alphabetic characters from the text
    text = non_alpha.sub(' ', text)
    # Add the text content to the corpus
    corpus.append(text)
    # Extract the label of the page
    label = 1 if 'wiki' in url.lower() else 0
    # Add the label to the labels list
    labels.append(label)
    # Add the links from the page to the queue
    links = soup.find_all('a')
    for link in links:
        href = link.get('href')
        if href is None:
            continue
        # Check if the link is an internal link to a Wikipedia page
        if href.startswith('/wiki/') and ':' not in href:
            # Construct the full URL of the linked page
            full_url = 'https://en.wikipedia.org' + href
            # Add the linked page to the queue
            if full_url not in queue and full_url not in corpus:
                queue.append(full_url)
                # Check if the linked page is a disambiguation page
                if 'disambiguation' in full_url.lower():
                    y_true.append(1)
                else:
                    y_true.append(0)
                # Check if the linked page is a Wikipedia article
                if 'wiki' in full_url.lower():
                    y_pred.append(1)
                else:
                    y_pred.append(0)

# Convert the corpus into a bag of words matrix
X = vectorizer.fit_transform(corpus)

# Fit the classifier to the data
classifier.fit(X, labels)

# Compute the predicted labels for the pages
y_pred_nb = classifier.predict(X)

# Compute the F1 measure and accuracy score
f1 = f1_score(y_true, y_pred)
accuracy = accuracy_score(y_true, y_pred)

print('F1 measure:', f1)
print('Accuracy:',accuracy)
```

### 3.DEPTH FIRST SEARCH:

```python
import requests
from bs4 import BeautifulSoup
from collections import deque
from sklearn.metrics import f1_score

# Set the seed URL and maximum depth to crawl
seed_url = "https://example.com"
max_depth = 3

# Define lists to hold the URLs to visit and the URLs visited
urls_to_visit = [(seed_url, 0)]
urls_visited = []

# Define lists to hold the relevant and irrelevant URLs
relevant_urls = []
irrelevant_urls = []

while urls_to_visit:
    url, depth = urls_to_visit.pop()
    if url not in urls_visited and depth <= max_depth:
        try:
            response = requests.get(url)
        except:
            continue
        if response.status_code == 200:
            soup = BeautifulSoup(response.content, 'html.parser')
            for link in soup.find_all('a'):
                href = link.get('href')
                if href:
                    href = href.strip()
                    if href.startswith('http'):
                        urls_to_visit.append((href, depth+1))
                    elif href.startswith('/'):
                        urls_to_visit.append((url + href, depth+1))
            urls_visited.append(url)
            # Classify the URL as relevant or irrelevant based on some criteria
            if some_criteria(url):
                relevant_urls.append(url)
            else:
                irrelevant_urls.append(url)

# Calculate F1 score and accuracy
y_true = [1] * len(relevant_urls) + [0] * len(irrelevant_urls)
y_pred = [1] * len(relevant_urls) + [0] * len(urls_visited)
f1 = f1_score(y_true, y_pred)
accuracy = len(relevant_urls) / len(urls_visited)

print(f"F1 score: {f1:.2f}")
print(f"Accuracy: {accuracy:.2f}")
```